\documentclass[12pt]{iopart}
\usepackage{iopams}
\usepackage[dvips]{graphicx}
\newcommand{\rhoB}{\rho_{\scriptscriptstyle B}}
\newcommand{\rhoC}{\rho_{\scriptscriptstyle C}}
\newcommand{\ssB}{{\scriptscriptstyle B}}
\newcommand{\Slash}[1]{\ooalign{\hfil/\hfil\crcr$#1$}}
\usepackage{color}

\newcommand{\Psfig}[2]{\includegraphics[width=#1]{PS/#2}}
\newcommand{\MeV}{\mathrm{MeV}}
\newcommand{\fm}{\mathrm{fm}}
\newcommand{\SU}{\mathrm{SU}}
\newcommand{\EOSY}{EOS$Y$}
\newcommand{\EOSYpi}{EOS$Y\pi$}
\begin{document}
\title[Tables of Hyperonic Matter Equation of State for Core-Collapse Supernovae]
{Tables of Hyperonic Matter Equation of State \\ for Core-Collapse Supernovae
\footnote{
{\tt http://nucl.sci.hokudai.ac.jp/$\widetilde{~}$chikako/EOS}
}}
\author{Chikako Ishizuka$^1$, Akira Ohnishi$^{1,2}$, Kohsuke Tsubakihara$^1$
Kohsuke Sumiyoshi$^{3}$ and Shoichi Yamada$^4$
}
\address{
$^1$ Department of Physics, Faculty of Science,\\
     Hokkaido University, Sapporo 060-0810, Japan
\\
$^2$ Yukawa Institute for Theoretical Physics, Kyoto University, Kyoto, Japan\\
$^3$ Numazu College of Technology, Numazu, Japan \\
$^4$ Science and Engineering, Waseda University, Tokyo, Japan
}
\eads{
\mailto{chikako@nucl.sci.hokudai.ac.jp}, 
\mailto{ohnishi@yukawa.kyoto-u.ac.jp}, 
\mailto{sumi@numazu-ct.ac.jp}
}

%
\begin{abstract}
We present sets of equation of state (EOS) of nuclear matter including hyperons
using an $\mathrm{SU}_f(3)$ extended relativistic mean field (RMF) model
with a wide coverage of density, temperature, and charge fraction
for numerical simulations of core collapse supernovae.
Coupling constants of $\Sigma$ and $\Xi$ hyperons with the $\sigma$ meson
are determined to fit the hyperon potential depths in nuclear matter,
$U_\Sigma(\rho_0) \simeq +30 \mathrm{MeV}$
and 
$U_\Xi(\rho_0) \simeq -15 \mathrm{MeV}$,
which are suggested from recent analyses of hyperon production reactions.
At low densities, the EOS of uniform matter is connected
with the EOS by Shen et al.,
in which formation of finite nuclei is included
in the Thomas-Fermi approximation.
In the present EOS, the maximum mass of neutron stars decreases
from $2.17 M_\odot$ ($Ne\mu$)
to $1.63 M_\odot$ ($NYe\mu$) when hyperons are included.
In a spherical, adiabatic collapse of a 15$M_\odot$ star by the hydrodynamics
without neutrino transfer,
hyperon effects are found to be small,
since the temperature and density do not reach the region of hyperon mixture,
where the hyperon fraction is above 1 \%
($T > 40 \mathrm{MeV}$ or $\rhoB > 0.4~\mathrm{fm}^{-3}$).
\end{abstract}

%
%
\section{Introduction}
\label{Sec:Intro}
The equation of state (EOS) plays an important role in high density 
phenomena
such as
high energy heavy-ion collisions,
neutron stars,
supernova explosions,
and black hole formations%
~\cite{Glendenning,Schaffner,Balberg,HYP00-Sahu,HYPMAT,Sugahara1994,Shen2002,PionCondensation,Dense,Sumiyoshi2004,Sumiyoshi2005,Sumiyoshi2006,Sumiyoshi2007,sQGP}.
The recent discovery of the strongly interacting
quark gluon plasma (sQGP)~\cite{sQGP} attracts attentions to
the EOS and transport coefficients
in the quark gluon plasma (QGP).
The core region of neutron stars,
where matter becomes very dense ($\sim 10^{15} \mathrm{g}/\mathrm{cm}^3$),
is an interesting play ground of quark and hadronic matter models.
Various ideas for the new form inside neutron stars have been proposed
including strangeness and quark
degrees of 
freedom~\cite{Glendenning,Schaffner,Balberg,HYP00-Sahu,HYPMAT,Sugahara1994,Shen2002,PionCondensation,Dense}.
Core-collapse supernovae also involve high density and temperature.
The nuclear repulsion at high densities drives the shock wave
at core bounce and the passage of shock wave heats up the matter
inside the supernova core.
A hot, lepton-rich neutron star (proto-neutron star)
is born after the explosion
and cools down by emitting supernova neutrinos.
When black holes are formed for more massive cores,
extremely high density and temperature are involved,
where hyperons should appear and quarks would be deconfined.
In order to describe the whole evolution of core-collapse supernovae
by numerical simulations,
one needs to prepare the set of microphysics under such extreme conditions.
One of the most important ingredients is the set of equation of state
(EOS) that contains necessary physical quantities.
It is to be noted that one must cover a wide range of
temperature, density and composition
in a consistent manner and theoretical framework.

Until now, the two sets of EOS
(Lattimer-Swesty EOS~\cite{LS-EOS} and Shen EOS~\cite{Shen-EOS})
have been widely used and applied
to numerical simulations of core-collapse supernovae~\cite{Sumiyoshi2004,Sumiyoshi2005}
and black hole formations~\cite{Sumiyoshi2006,Sumiyoshi2007}.
The Lattimer-Swesty EOS is based on a compressible liquid-drop model,
whose mass and mean field potential are motivated
by non-relativistic zero-range Skyrme type interactions.
The Shen EOS is based on a relativistic mean field (RMF) model,
whose interactions are determined by fitting the binding energies
and nuclear radii of stable as well as unstable nuclei~\cite{TM1}.
Coexistence of nuclei and uniform matter is included
in the Thomas-Fermi approximation in the Wigner-Seitz cell,
and the alpha particles are assumed to follow the statistical
distribution with excluded volume effects.

The constituents in these EOSs are neutrons, protons,
alpha-particles and nuclei, restricting the framework
within the non-strange baryons.
These degrees of freedom may be enough to simulate the
early stage of hydrodynamical evolution of supernova explosions.
However, in order to clarify
the long-time evolution from core collapse~\cite{Sumiyoshi2005}
to proto-neutron star cooling~\cite{Suzuki,Pons},
black hole formation~\cite{Sumiyoshi2006,Sumiyoshi2007},
neutron star mergers
and gamma ray bursts
that may involve higher density/temperature,
it would be necessary to include other particle degrees of freedom.
Especially, hyperons (baryons containing strange quarks) are
commonly believed to appear in neutron star core and to modify
the neutron star profile~\cite{Glendenning,Schaffner,Balberg,HYP00-Sahu,HYPMAT,Sugahara1994,Shen2002}.
While there are several works which include the hyperons
in the proto-neutron star cooling~\cite{Pons},
there has been no study on the dynamics of core-collapse supernovae
adopting the EOS with hyperons.
This is partially because EOS table of supernova matter
including hyperons has not been available in public.
In addition, the determination of
the interaction for hyperons has been difficult
having large uncertainties so far.

Recently, developments in hypernuclear physics
have narrowed down the allowed range of hyperon potential depth
in nuclear matter.
The potential depth of $\Lambda$ has been well known to be
around $U^{(N)}_\Lambda(\rho_0) \simeq -30 \mathrm{MeV}$
from bound state spectroscopy.
For $\Sigma$ baryons,
it was considered to feel similar potential to $\Lambda$,
because it contains the same number of light $(u,d)$ quarks.
From the recently observed quasi-free $\Sigma$ production spectra~\cite{Noumi},
it is now believed that $\Sigma$ baryons would feel repulsive potential
in nuclear matter; $U_\Sigma(\rho_0) \simeq +30 \mathrm{MeV}$%
~\cite{Harada,Kohno,Maekawa},
Also for $\Xi$ baryons, the analyses of
the twin hypernuclear formation~\cite{Twin}
and the $\Xi$ production spectra~\cite{Fukuda,HYP06-Maekawa,MaekawaXi},
suggest the potential depth of around
$U^{(N)}_\Xi(\rho_0) \simeq -15 \mathrm{MeV}$.
These $\Sigma$ and $\Xi$ hyperons are particularly important in neutron stars,
since nuclear matter can take a large energy gain
from neutron Fermi energy and symmetry energy by replacing, 
for example, two neutrons
with a proton and a negatively charged hyperon ($\Sigma^-$ or $\Xi^-$).
The updates on the interactions of hyperons may have impact on supernova
dynamics and thermal evolution of proto-neutron stars.

In this paper, we present new sets of EOS of dense matter
with hyperons, abbreviated as \EOSY,
under the current understanding of interaction.
We provide the data table
covering a wide range of temperature ($T$), density ($\rhoB$),
and charge-to-baryon number ratio in hadronic part ($Y_C$),
which enables one to apply to supernova simulations.
Our framework is based on
the RMF theory with the parameter set TM1~\cite{TM1},
which was used to derive the EOS table by Shen et al.~\cite{Shen-EOS},
and is extended to include hyperons by considering 
the flavor SU(3) Lagrangian~\cite{Schaffner}.
Therefore, our EOS table is smoothly connected with Shen EOS
and can be used easily as an extension
of Shen EOS table in numerical simulations.

It is well known that the RMF predicts
large values of incompressibility ($K \sim 300~\mathrm{MeV}$)
and symmetry energy,
and these are sometimes considered to cause problems in applying
to dense matter EOS, since they lead to too high maximum mass of neutron stars
without hyperons and may be unfavorable to core-collapse explosions.
It should be noted that $K$ and symmetry energy values are not yet well
determined separately in a model independent manner.
Analyses of collective flow data at AGS energies suggest
$K=210-300~\mathrm{MeV}$~\cite{Danielewicz,Sahu2000},
and collective flows at SPS energies are shown 
to be more sensitive to the mean field of resonance hadrons
rather than to the cold matter EOS~\cite{Isse}.
For symmetry energy, it is possible to describe
binding energies, proton-neutron radius differences and 
isovector giant monopole resonances simultaneously
by incorporating density dependent coupling (DD-ME1)
in RMF and relativistic RPA~\cite{RRPA} with a larger value of
symmetry energy than that in non-relativistic models.
Note that the interaction of RMF-TM1 is constrained
by the nuclear masses, radii, neutron skins and excitations~\cite{TM1}.
The large value of $K$ leads to a large neutron star mass (stiff EOS),
which seems not preferable for explosion.
On the other hand, the large symmetry energy is known to
be preferable for explosion, having less free proton fraction
and smaller electron captures. These two effects are competing
each other in the sophisticated numerical simulations~\cite{Sumiyoshi2005}.
We note also that the RMF fulfills automatically the causality
(the sound velocity should not exceed the light velocity)
whereas the non-relativistic frameworks breaks down at high densities
appearing in the simulations.
Thus at present we do not find problems in applying RMF EOS to dense matter
compared to non-relativistic models.

This paper is arranged as follows.
In section~\ref{Sec:Model},
we describe the framework to calculate the
dense matter at finite temperature including hyperons.
We explain the updated information on
hyperon potentials in nuclear matter
and adopted potential values in \EOSY.
We describe also the prescriptions to provide the data
table for the wide range of density
including sub-saturation densities where finite nuclei appear.
In section~\ref{Sec:Results},
we report the properties of \EOSY\ 
in comparisons with nucleonic EOS (TM1/Shen EOS).
We apply \EOSY\ to cold neutron star matter and supernova matter.
We show several properties of \EOSY\ at finite temperatures
by examining energies, chemical potentials and compositions.
The data tables are successfully applied to hydrodynamical calculations
of adiabatic collapse of iron core of massive stars.
We examine the possibilities of hyperon appearance in supernova cores.
Summary and discussions are given in section~\ref{Sec:Summary}.
In the appendix, we provide the the definitions of quantities
in \EOSY.

\section{Model and Method}
\label{Sec:Model}

In this work, we construct the EOS table of supernova matter
based on a relativistic mean field model.
We adopt the parameter set TM1~\cite{TM1} for non-strange sector.
For its flavor SU(3) extension,
we start from the work by Schaffner and Mishustin~\cite{Schaffner},
and we include the updated information on hyperon potentials
from recent experimental and theoretical hypernuclear physics developments.
Low density part of the EOS is connected with the Shen EOS.

\subsection{Relativistic mean field model with hyperons}
\label{Subsec:RMFY}

The relativistic mean field (RMF) theory is constructed to describe
nuclear matter and nuclei based
on the relativistic Br{\"u}ckner-Hartree-Fock theory~\cite{RBHF},
which successfully describes the nuclear matter saturation.
It is preferable to adopt the relativistic frameworks
for astrophysical applications,
since they automatically satisfy the causality,
{\it i.e.} the sound velocity is always less than the speed of light.

The RMF parameter set TM1 is determined to describe
binding energies and nuclear radii
of finite nuclei from Ca to Pb isotopes
and fulfills the nuclear matter saturation.
The incompressibility of symmetric uniform matter
and the symmetry energy parameters are found to be
$K=281~\mathrm{MeV}$ and $a_{sym}=36.9~\mathrm{MeV}$.
When it is applied to neutron stars,
the maximum mass of cold neutron stars with TM1 is $2.17 M_\odot$.

The extension of the RMF to flavor SU(3) has been investigated by many authors.
A typical form of the Lagrangian density including hyperons
is given as~\cite{Schaffner},
\begin{eqnarray}
&&{\cal L}
= \sum_B\bar{\Psi}_B \left(i\Slash{\partial} - M_B\right)\Psi_B
+\frac{1}{2}\partial^\mu\sigma\partial_\mu\sigma-U_\sigma(\sigma)
\nonumber\\
&&~~
 -\frac{1}{4}\omega^{\mu\nu}\omega_{\mu\nu}
  +\frac{1}{2}m_\omega^2\omega^\mu \omega_\mu 
  -\frac{1}{4}\vec{R}^{\mu\nu}\cdot\vec{R}_{\mu\nu}
  +\frac{1}{2}m_\rho^2\vec{R}^\mu\cdot\vec{R}_\mu
\nonumber\\
&&~~
 -\sum_B \bar{\Psi}_B\left(
	 g_{\sigma B}\sigma
	+g_{\omega B}\Slash{\omega}
	+g_{\rho B}\vec{\Slash{R}}\cdot\vec{t}_B
	\right)\Psi_B
  +\frac14\,c_\omega(\omega^\mu\omega_\mu)^2
+{\cal L}^{YY} \ , \nonumber \\
&&U_\sigma(\sigma) = \frac{1}{2}m_\sigma^2\sigma^2 +\frac{g_3}{3}\sigma^3
+\frac{g_4}{4}\sigma^4 \ , \nonumber \\
&&{\cal L}^{YY} =
\frac12\partial_\nu\zeta\partial^\nu\zeta
-\frac12m_\zeta^2\zeta^2
-\frac{1}{4}\phi_{\mu\nu}\phi^{\mu\nu}
+ \frac{1}{2}m_\phi^2\phi_\mu\phi^\mu 
\nonumber \\
&&~~~~
- \sum_B \bar{\Psi}_B\left(g_{\zeta B}\zeta+g_{\phi B}\gamma^\mu\phi_\mu\right)
	\Psi_B
\ ,
\end{eqnarray}
where the sum runs over all the octet baryons.
In this Lagrangian, hidden strangeness ($\bar{s}s$) scalar and vector
mesons, $\zeta$ and $\phi$, are included in addition to 
$\sigma$, $\omega$ and $\rho$ (represented by $\vec{R}^\mu$) mesons.
Strength tensors of $\omega$, $\rho$ and $\phi$ mesons are shown
in $\omega^{\mu\nu}$, $\vec{R}^{\mu\nu}$ and $\phi^{\mu\nu}$, respectively.
The Lagrangian contains meson masses, coupling constants,
and self-coupling constants as parameters.

In introducing hyperons in RMF,
we have large ambiguities in hyperon-meson coupling constants.
One of the ways to determine the parameters is to rely on symmetries.
Schaffner and Mishustin~\cite{Schaffner} have determined
hyperon-vector meson coupling constants
based on the $\mathrm{SU}(6)$ (flavor-spin) symmetry,
\begin{eqnarray}
\frac13 g_{\omega N}
=\frac12 g_{\omega\Lambda}
=\frac12 g_{\omega\Sigma}
=g_{\omega\Xi}
\ ,
g_{\rho N}=\frac12 g_{\rho\Sigma}=g_{\rho\Xi}\ ,\quad g_{\rho\Lambda}=0\ ,
\\
2g_{\phi\Lambda}=2g_{\phi\Sigma}=g_{\phi\Xi}=-\frac{2\sqrt{2}}{3}g_{\omega N}
\ ,\quad 
g_{\phi N}=0\ .
\end{eqnarray}
Scalar mesons in RMF may partially represent contributions from 
some other components than $\bar{q}q$, such as $\pi\pi$ in $\sigma$.
In Ref.~\cite{Schaffner}, 
the scalar meson couplings to hyperons have been given
based on the assumption that hyperons feel
potentials in nuclear and hyperon matter as,
\begin{eqnarray*}
U_\Lambda^{(N)} = U_\Sigma^{(N)} = -30~\mathrm{MeV}\,,  \quad
U_\Xi^{(N)} =~-28 \mathrm{MeV}\,, \\
U_\Sigma^{(\Sigma)}\sim U_\Lambda^{(\Sigma)}
\sim U_\Sigma^{(\Lambda)}\sim 2U_\Lambda^{(\Lambda)}\sim -40~\mathrm{MeV}\,,
\end{eqnarray*}
where $U^{(B')}_B$ denotes the potential of $B$ in baryonic matter
at around $\rho_0$ composed of $B'$.
Recent developments in hypernuclear physics
suggest that hyperon potentials in nuclear matter
are repulsive for $\Sigma$~\cite{Noumi,Harada,Kohno,Maekawa},
and weakly attractive for $\Xi$~\cite{Twin,Fukuda,HYP06-Maekawa,MaekawaXi},
respectively.

$\Xi$ hyperons are expected to have nuclear bound states,
and the bound state spectroscopy at forthcoming facilities such as
J-PARC and FAIR will give a strong constraints
on the $\Xi$ potential in nuclear matter.
At present, the depth of the $\Xi^-$-nucleus potential
has been suggested to be around 15 MeV from the analysis of
twin hypernuclear formation~\cite{Twin}
and the $(K^-,K^+)$ spectrum in the bound state region~\cite{Fukuda}.
In the former,
the binding energy of the $\Xi^-$-nuclear system is found to be
consistent with a shallow $\Xi^-$-nuclear potential
in an event accompanied by two single hyperfragments
emitted from a $\Xi^-$ nuclear capture at rest (a twin hypernuclei)
found in a nuclear emulsion~\cite{Twin}.
In the latter,
while the resolution of experimental data is not enough to distinguish 
the bound state peaks,
the observed yield or the spectrum shape in the bound state region 
is found to be in agreement with the calculated results with 
$U_\Xi^{(N)} \simeq -15~\MeV$~\cite{Fukuda,HYP06-Maekawa,MaekawaXi}.

For $\Sigma$ hyperons, it is necessary to analyze continuum spectra.
In the observed (quasi-)bound $\Sigma$ nucleus ${}^4_\Sigma$He~\cite{He4Sig},
the coupling effect is strong and the repulsive contribution
in the $T=3/2$, $^3S_1$ channel is suppressed,
then it does not strongly constrain the $\Sigma$ potential in nuclear matter.
The analysis of $\Sigma^-$ atomic data suggested
a $\Sigma^-$-nucleus potential having a shallow attractive pocket
around the nuclear surface and repulsion inside the nucleus~\cite{Batty,Mares,HYP06-Tsubaki},
but the atomic energy shift is not sensitive to the potential
inside the nucleus.
In the distorted wave impulse approximation (DWIA) analyses
of the quasi free (QF) spectrum
in the continuum region~\cite{Noumi,Harada,Kohno,Maekawa},
it is suggested that the $\Sigma$ hyperon would feel repulsive real potential
of $10\sim90 \mathrm{MeV}$.
Recent theoretical analyses favor the strength of repulsion
of around $+30 \mathrm{MeV}$~\cite{Harada,Kohno,Maekawa}.
This repulsion may come from the Pauli blocking effects between quarks
due to the isovector nature of the diquark pair in $\Sigma$~\cite{KohnoQC}.
In a Quark-Meson Coupling (QMC) model, medium modification 
of the color hyperfine interaction in the quark bag is found to be
the origin of repulsive $\Sigma$ potential~\cite{Stone}.
The $\Sigma$ potential in nuclear matter at saturation density
is predicted to be around $+30$ MeV (repulsion)
in a quark cluster model $YN$ potential \cite{KohnoQC},
and a chiral model also predicts a similar repulsion~\cite{Kaiser}.

From these discussions, we adopt the following potential strength
as {\em recommended} values,
\begin{equation}
U_\Sigma^{(N)} (\rho_0) \simeq +30~\mathrm{MeV}\,,
\quad
U_\Xi^{(N)} (\rho_0) \simeq -15~\mathrm{MeV}  \ .
\label{Eq:HypPot}
\end{equation}

The above spectroscopic studies have been done mainly with non-relativistic
frameworks for hyperons,
then the potential should be regarded
as the Schr\"odinger equivalent potential in RMF.
The Schr{\"o}dinger equivalent potential
is related to the scalar ($U_s$) and vector ($U_v$) potentials as,
\begin{eqnarray}
U_\ssB(\rho, E(\bold{p}))
&=&U_s(\rho)+{E(\bold{p})\over M_\ssB}\,U_v(\rho)
\nonumber\\
&=& g_{\sigma B}\sigma
 +  g_{\zeta B}\zeta
 + {E\over M}\,\left(
    g_{\omega B}\omega
 +  g_{\rho B} R
 +  g_{\phi B}\phi
 \right)
\ ,
\end{eqnarray}
where $R$ represents the expectation value of the $\rho$ meson.
We have fixed $g_{\sigma B}$ value
by fitting the hyperon potential depth in normal symmetric nuclear matter,
\begin{equation}
U_\ssB^{(N)}(\rho_0)
= g_{\sigma B}\sigma^{(N)}(\rho_0)
+ g_{\omega B}\omega^{(N)}(\rho_0)
\ ,
\end{equation}
where $\sigma^{(N)}(\rho_0)$ and $\omega^{(N)}(\rho_0)$ represent
the expectation values of $\sigma$ and $\omega$ mesons
in symmetric nuclear matter at $\rho_0$.
We adopt the parameter set TM1 for nucleon sector,
and we determine $g_{\sigma\Sigma}$ and $g_{\sigma\Xi}$
to reproduce the potential depths
of $\Sigma$ and $\Xi$ hyperons in Eq.~(\ref{Eq:HypPot})
as listed in Table~\ref{Table:TM1}.
We choose other hyperon-meson coupling constants referring to the values
in Ref.~\cite{Schaffner}.
We show the values of $g_{\sigma\Sigma}$ and $g_{\sigma\Xi}$ 
for different potentials in Table~\ref{Table:Potential}.

\begin{table}
\caption{The coupling constants of the parameter sets.}
\label{Table:TM1}
\begin{center}
\begin{tabular}{cccc} \hline
$m_\sigma$ (MeV) & $g_3$ (MeV) & $g_4$ & $c_\omega$
\\\hline
511.198 & 1426.466 & 0.6183 & 71.3075
\\\hline
\end{tabular}
\\[1ex]
\begin{tabular}{c|ccccc} \hline
$g_{MB}$  & $\sigma$	& $\zeta$& $\omega$	&$\rho$		&$\phi$\\\hline
$N$	  & 10.0289	& 0	 & 12.6139	&4.6322		& 0	\\
$\Lambda$ & 6.21	& 6.67   & 8.41		&0		&$-5.95$\\
$\Sigma$  & 4.36	& 6.67   & 8.41		&9.26		& $-5.95$\\
$\Xi$	  & 3.11	& 12.35  & 4.20		&4.63		&$-11.89$\\
\hline
\end{tabular}
\end{center}
\end{table}

\begin{table}
\caption{The coupling constants of $\Sigma{N}$ and $\Xi{N}$.}
\label{Table:Potential}
\begin{center}
\begin{tabular}{ccc} \hline
$U^{(N)}_\Sigma(\rho_0)$ (MeV) & $g_{\sigma\Sigma}$ \\
\hline
$+90$& $2.58$ & \\ 
$+30$& $4.36$ & present \\ 
$  0$& $5.35$ & \\ 
$-10$& $5.63$ & \\ 
$-30$& $6.21$ & Ref.~\cite{Schaffner} \\
\hline
$U^{(N)}_\Xi(\rho_0)$ (MeV) & $g_{\sigma\Xi}$ \\
$-15$ & $3.11$ & present \\ 
$-28$ & $3.49$ & Ref.~\cite{Schaffner} \\ \hline
\end{tabular}
\end{center}
\end{table}

In the literatiure, the instatibility due to the negative effective mass
of nucleon has been reported~\cite{Knorren,Menezes}.
As $\sigma$ increases,
the nucleon effective mass reaches zero at $\sigma=M_N/g_{\sigma N}$
where hyperon effective masses are still positive and
act to further increase $\sigma$,
leading to the nucleon negative effective mass.
This effect depends very much on $\sigma{Y}$ couplings,
which are small within the current sets of parameters,
and we did not find this instability in the range of data tables.
However, we found this instability occurs at very high densities/temperatures
which are relevant in the black hole formation~\cite{SumiyoshiNext}.

\subsection{Free thermal pions}
\label{Subsec:pions}

In black hole formation processes as found in Ref.~\cite{Sumiyoshi2006},
the temperature goes up to around 100 MeV.
At these high temperatures, pion contributions become dominant.
Charged pions may condensate at high densities
in neutron star matter~\cite{Glendenning,PionCondensation}.
To estimate the effect of pion mixtures,
we also prepare the EOS table including free thermal pions
assuming the pion mass is not affected by the interaction.
This is of course oversimplification, however, the first trial to include pions.
Further sophisticated studies are necessary.

The density of free thermal pions is calculated to be
\begin{equation}
\rho_\pi = \rho_\pi^{Cond}
+ \int {d^3p \over (2\pi)^3}\,{1 \over \exp((E_\pi(\bold{p})-\mu_\pi)/T)-1}
\ ,
\end{equation}
where $\mu_\pi=\mu_C, 0, -\mu_C$ for $\pi^+, \pi^0, \pi^-$, respectively.
When the absolute value of the chemical potential reaches the pion mass,
pion condensation occurs;
{\em i.e.} 
the amount of condensed $\pi$ at zero momentum can take any value
at $\mu_C=\pm m_\pi$.
We have determined the amount of condensed $\pi$ in the following way.
First we solve the equilibrium condition and obtain $\mu_B$ and $\mu_C$
without condensed $\pi$.
When $|\mu_C| > m_\pi$, we set $|\mu_C|=m_\pi$ and re-evaluate hadron 
densities except for the condensed $\pi$ to satisfy the condition
of $\rhoB = \rhoB(\mathrm{Given})$.
Finally, the amount of condensed $\pi$ is given so as to satisfy
the charge density condition, $\rhoC=Y_C(\mathrm{Given})\rhoB$.

The pion condensation in the current treatment
is a simple $s$-wave Bose-Einstein condensation,
which is different from the pion condensation derived from $p$-wave $\pi N$
interaction~\cite{PionCondensation}.
We mention that pion condensation will be suppressed
after considering $s$-wave $\pi N$
repulsive interaction discussed in the energy shift of
deeply bound pionic atoms~\cite{PionicAtom,Batty1983,Kienle}
and in pion-nucleus scattering~\cite{Friedman}.

\subsection{Low density}
By using these potentials, 
we can immediately obtain a EOS of uniform dense matter with strangeness
based on the RMF theory.
We also need to cover the low-density region below $\rho_0$,
where the inhomogeneous matter appears.
Here we connect the uniform matter EOS with Shen EOS~\cite{Shen-EOS},
which is based on the same RMF parameter set TM1
and treats the inhomogeneity with the Thomas-Fermi approach.

We include the contribution from inhomogeneity
by adding the free energy difference of Shen
EOS values from those in uniform matter,
\begin{eqnarray}
F = F_{RMF}^Y + \Delta F_{Nucl}
\end{eqnarray}
at $\rhoB \le \rho_0$, where
\begin{eqnarray}
\Delta F_{Nucl} = F_{Shen}- F_{RMF}^{(np)} \ .
\end{eqnarray}
Other variables are derived in the same way as the above equations.
The deviation due to inhomogeneity $\Delta F$ vanishes at $\rhoB > \rho_0$.
These prescriptions produce the extended EOS tables for studies in astrophysics,
containing inhomogeneity at low density and strangeness information.
The compositions of $n, p, \alpha, A, Y$ are consistent with Shen EOS table
and the sum of each component ratio becomes unity.

\subsection{Tabulation of thermodynamical quantities}
Thermodynamical quantities are provided in the data table 
as a function of 
baryon mass density $\rhoB$, charge ratio $Y_C$, 
and temperature $T$.  
Here $Y_C$ means the charge ratio defined as $Y_C = n_C / n_B$ 
and $n_C$ is a charge density.
See Appendix for the list of quantities and their definitions,
which are slightly revised from the original table of Shen EOS.

For the purpose of numerical simulations, we prepare the EOS table 
containing the contributions of leptons and photons
by adding the energy, pressure and entropy from 
electrons, positrons and photons to the hadronic EOS.
We treat electrons and positrons as ideal Fermi gas with the finite 
rest mass and calculate photons according to the standard expressions 
for radiations.

The baryon mass density, charge ratio and temperature 
cover the following range,
\begin{center}
\begin{itemize}
\item $\rhoB$ = $10^{5.1}$ $\sim$ $10^{15.4}$ ($g/\mathrm{cm}^{3}$)
(104 points)
\item $Y_C = 0$ and $0.01 \sim 0.56$
(72 points)
\item $T = 0$ and $0.1 \sim 100$ (MeV)
(32 points)
\end{itemize}
\end{center}
Mesh points for $\rhoB$, $Y_C(>0)$ and $T(>0)$
are taken as approximate geometric sequences with
$\Delta\log_{10}\rhoB=0.1$, 
$\Delta\log_{10}Y_C=0.025$ ($-2.00\leq\log_{10}Y_C\leq-0.25$)
and 
$\Delta\log_{10}T \simeq 0.1$,
respectively.
These ranges and mesh points are the same as those in Shen EOS.
By connecting smoothly in the way described above, 
the EOS table is combined with Shen EOS at lower densities below $\rho_0$ 
while it includes full baryon octet at high densities 
so that one can see the effects of hyperon mixture.  
Some tabulated quantities in the EOS table need attentions.
The values of the mass $A$ and charge number $Z$ of heavy nucleus 
are taken from Shen EOS at densities below $\rho_0$ 
and are set to be zero above $\rho_0$.  
Similarly, the fraction of $\alpha$-particle and heavy nucleus 
are taken from Shen EOS at low densities 
and are set to be zero at high densities.  

\section{Properties of EOS tables and astrophysical applications}
\label{Sec:Results}

We report the properties of dense matter in the present EOS table 
with hyperons (\EOSY)
and their applications to neutron stars and supernovae. 
We adopt hereafter the case of
$(U^{(N)}_\Sigma(\rho_0),U^{(N)}_\Xi(\rho_0))=(+30~\MeV, -15~\MeV)$
as a standard case,
which is currently the most recommended set of hyperon potentials.
We also consider
the case with pion contribution (EOS$Y\pi$)
and the attractive hyperon potential case~\cite{Schaffner}
$(U^{(N)}_\Sigma(\rho_0),U^{(N)}_\Xi(\rho_0))=(-30~\MeV, -28~\MeV)$,
abbreviated as EOS$Y$(SM).

\begin{table}[bth]
\caption{
Constituents, assumed hyperon potentials, threshold densities,
maximum masses of neutron stars,
central densities giving maximum masses of neutron stars
and neutron star masses at the threshold central densities
in nucleonic EOS (TM1/Shen EOS~{\protect\cite{TM1,Shen-EOS}}),
hyperonic EOS with attractive hyperon potentials
(\EOSY(SM)~{\protect\cite{Schaffner}}),
hyperonic EOS with repulsive hyperon potentials (\EOSY, present work)
and 
hyperonic EOS with repulsive hyperon potentials including pions
(\EOSYpi, present work).
For $\pi^-$, the maximum density of condensation is also shown in parentheses.
Threshold densities of protons and muons are
$1.1\times{10}^{-4}~\fm^{-3}$ and $0.11~\fm^{-3}$, respectively.
}\label{Table:NS}
\begin{center}
\begin{tabular}{c|cccc}
\hline
\hline
EOS	& TM1/Shen EOS	& EOS$Y$(SM)	& EOS$Y$	& EOS$Y\pi$	\\
\hline
Constituents& $Ne(\mu)$	& $NYe(\mu)$	& $NYe(\mu)$	& $NY\pi{e}(\mu)$\\
\hline
$U^{(N)}_\Sigma$ (MeV)& & $-30$		& $+30$		& $+30$		\\
$U^{(N)}_\Xi$ (MeV)   & & $-28$		& $-15$		& $-15$		\\
\hline
$\rho_{(thr)}(\fm^{-3})$\\
$\Lambda$ &		& 0.32		& 0.32		& 0.37		\\
$\Sigma^-$ &		& 0.29		& 1.14		& 1.1		\\
$\Sigma^0$ &		& 0.57		& 1.34		& 1.3		\\
$\Sigma^+$ &		& 0.69		& 1.47		& 1.5		\\
$\Xi^-$		&	& 0.43		& 0.40		& 0.56		\\
$\Xi^0$		&	& 0.62		& 0.71		& 0.74		\\
$\pi^-$		&	&		&		& 0.16(0.88)	\\
\hline
$M^{(max)}_{NS}(M_\odot)$
	& 2.17		& 1.55		& 1.63		& 1.65		\\
$\rhoB^{(max)}(\fm^{-3})$
	& 1.12		& 0.79		& 0.79		& 0.97		\\
$M^{(thr)}_{NS}(M_\odot)$
	&		&1.17($\Sigma^-$)&1.28($\Lambda$)& 1.22($\Lambda$)\\
	&		&		&		&0.51($\pi^-$)\\
\hline
\end{tabular}
\end{center}
\end{table}

\begin{figure}
\centerline{
\Psfig{16cm}{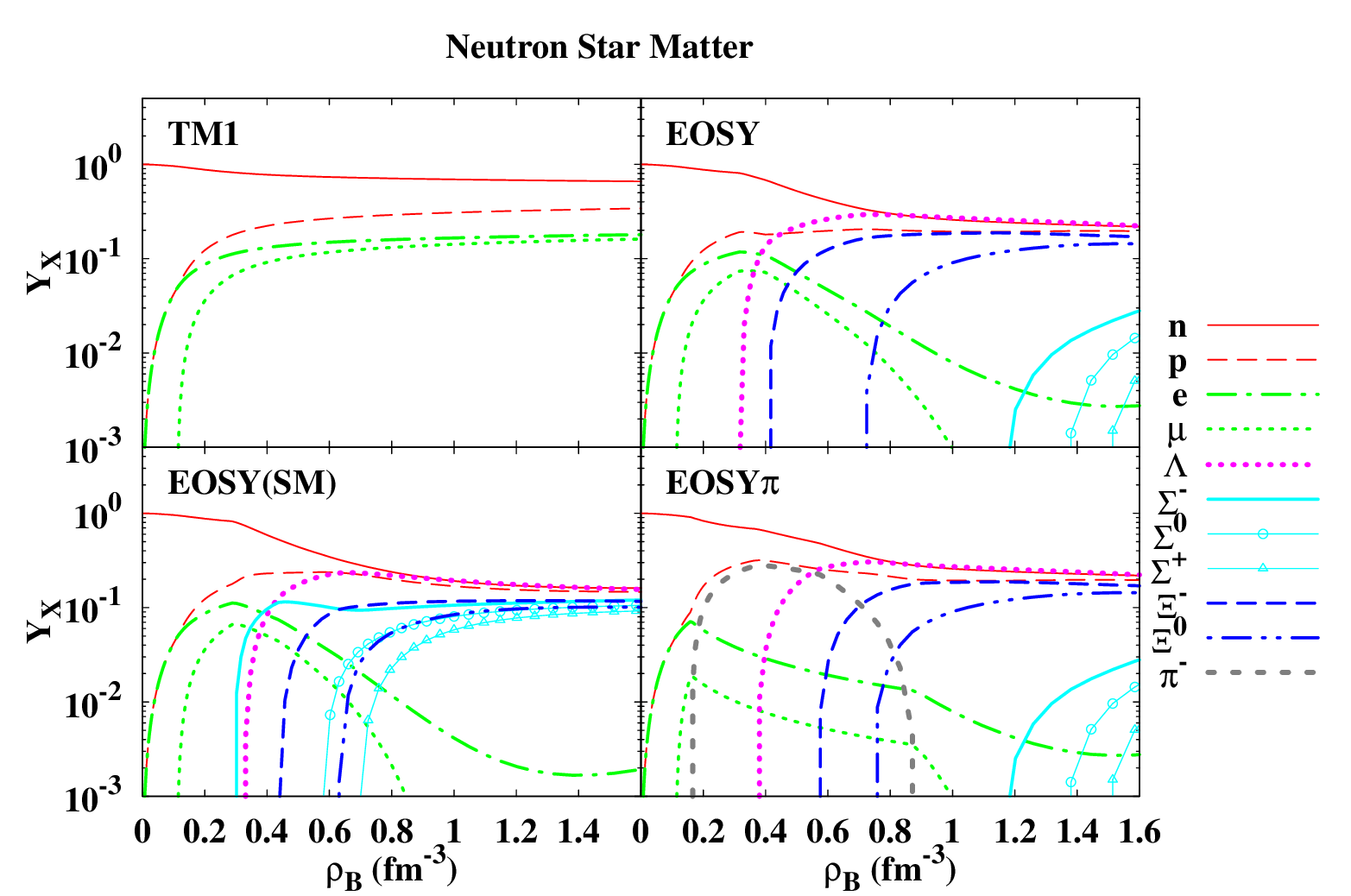}
}
\caption{Composition of neutron star matter
in nucleonic EOS (TM1, upper-left),
hyperonic EOS with attractive potential (\EOSY(SM), lower-left),
hyperonic EOS with repulsive potential without (\EOSY, upper-right)
and with pions (\EOSYpi, lower-right).
The number fraction of particles are plotted as functions 
of baryon density.  The species of particles are denoted 
as in the legend.}
\label{Fig:Const-NS}
\end{figure}

\subsection{Neutron star matter}

We first study the EOS of neutron star matter,
which is under the $\beta$ equilibrium at zero temperature.  
We here add electron and muon contributions under the $\beta$ equilibrium 
and charge neutrality conditions.
We consider uniform matter ignoring finite nuclear effects.

We show particle compositions in neutron star matter
in Fig.~\ref{Fig:Const-NS} 
to see the appearance of new degrees of freedom.  
We display the cases of
nucleonic (TM1, upper-left) and hyperonic (EOS$Y$, upper-right) EOS.
Results with hyperonic EOS with attractive $\Sigma$ potential
(EOS$Y$(SM), lower-left)
and hyperonic EOS with pions (EOS$Y\pi$, lower-right)
are also shown for comparison.
The particle composition of neutron star matter is very sensitive
to the choice of hyperon potentials.
With attractive $\Sigma$ potential,
$\Sigma^-$ appears at lower densities than $\Lambda$.
With repulsive $\Sigma$ potential,
$\Lambda$ appears first followed by $\Xi^-$ and $\Xi^0$.
This behavior is different from the previous works that adopt attractive
potentials~\cite{Glendenning,Schaffner,Balberg,HYP00-Sahu,HYPMAT},
and pointed out in Refs.~\cite{Schaffner,Balberg,HYP00-Sahu,Stone}.
When we allow the appearance of pions,
condensed pions ($\pi^-$) appear prior to hyperons.
With $\pi^c$ condensation,
the charge chemical potential is restricted to be $|\mu^c| \leq m_{\pi}$
and the proton fraction becomes larger,
then the neutron chemical potential is reduced.
As a result, the threshold density of hyperons are shifted up. 
The density of $\Lambda$ appearance is about $0.37~\fm^{-3}$
and other hyperons such as $\Xi^-$ are also suppressed.

\begin{figure}
\centerline{
\Psfig{16cm}{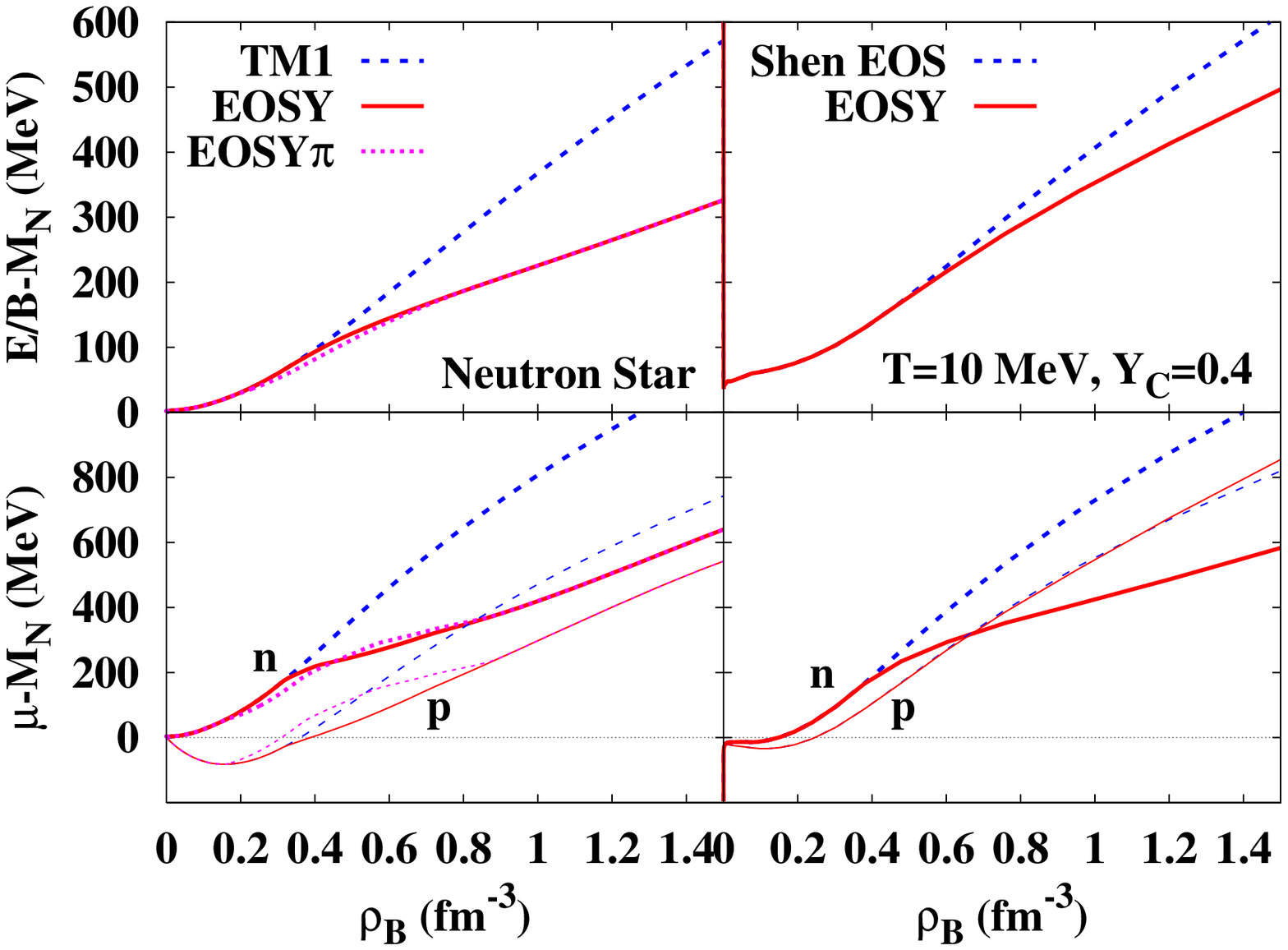}
}
\caption{EOS of neutron star matter and supernova matter
for nucleonic EOS (TM1/Shen EOS), hyperon EOS(\EOSY),
hyperon EOS with free pions (\EOSYpi).
The upper panels display the energy per baryon,
and the lower panels the chemical potentials of proton and neutron.
The dashed line is for the nucleonic EOS, the solid line is for 
hyperon EOS (\EOSY), and dotted line is for hyperon EOS with pions (\EOSYpi).
}
\label{Fig:E/B}
\end{figure}

\begin{figure}
\centerline{
\Psfig{10cm}{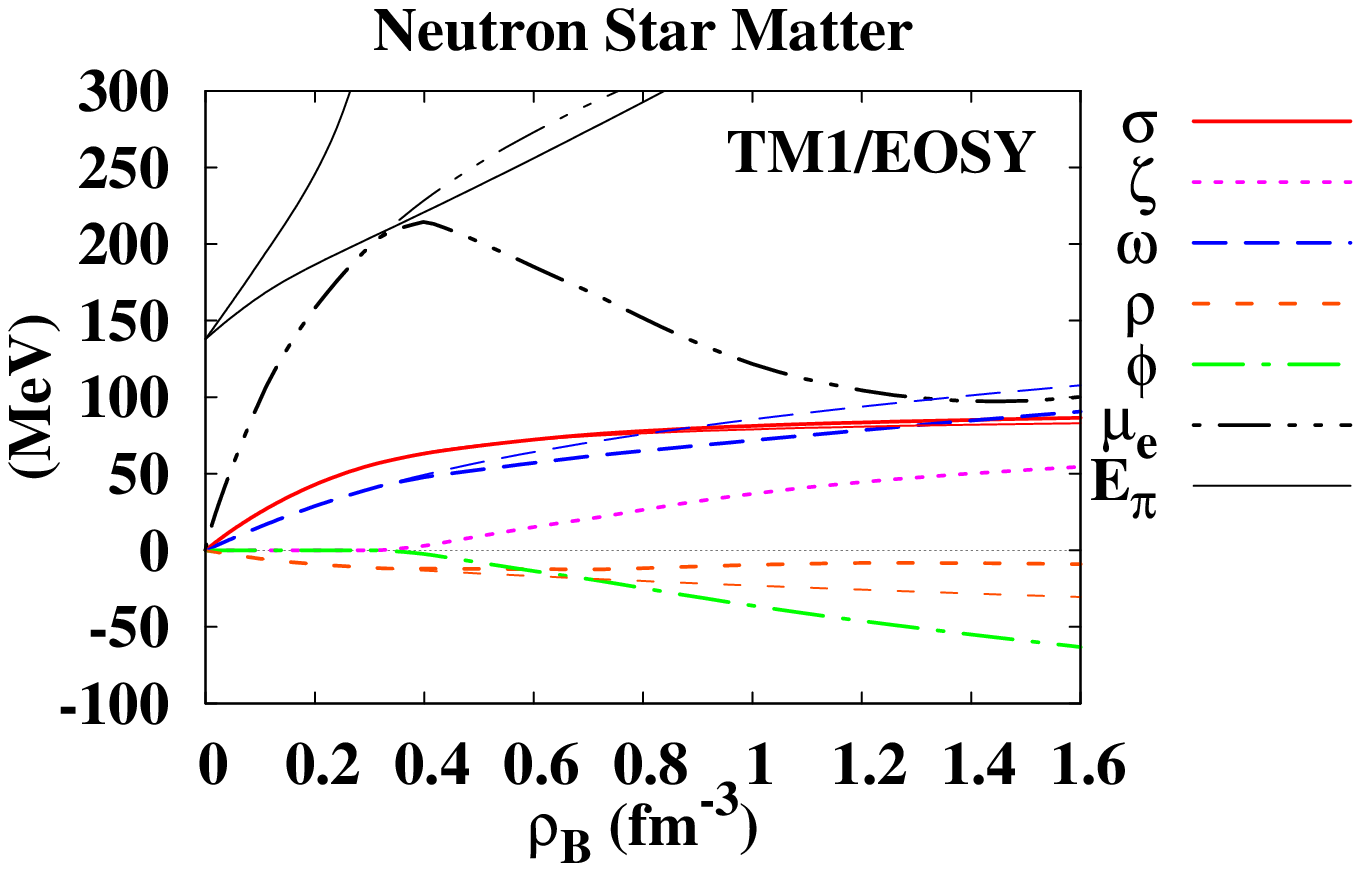}
}
\caption{Meson expectation values and electron chemical potential
in neutron star matter for TM1 EOS (thin lines)
and hyperon EOS(\EOSY) (thick lines).
Thin solid lines show the pion energy,
$E_\pi=\sqrt{m_\pi^2+2m_{\pi}U_s(\pi^-)}$,
where lower less repulsive (upper more repulsive) line shows the results
with the pion optical potential in Ref.~\protect{\cite{Batty1983}}
(\protect{\cite{Kienle}}) by using the proton fraction in TM1 EOS.
}
\label{Fig:Pot}
\end{figure}

In the left panel of Fig.~\ref{Fig:E/B},
we show the energy per baryon ($E/B$)
and chemical potentials ($\mu_n$ and $\mu_p$) in neutron star matter
in nucleonic EOS (TM1), \EOSY\,  and \EOSYpi.
Compared with nucleonic EOS,
$E/B$ is much lower in \EOSY\,  at high densities.
Chemical potentials are also suppressed with hyperons correspondingly.
There are several origins for this energy gain.
First, nucleon Fermi energy decreases due to the hyperon mixture.
Secondly, the lepton contribution is suppressed
when negatively charged hyperons emerge.
In addition, the repulsive vector potential becomes small, 
because the $\omega{Y}$ couplings are smaller than $\omega N$
and the isospin asymmetry becomes smaller
when negatively charged hyperons appear,
as shown in the long-dashed lines in Fig.~\ref{Fig:Pot}.

For $E/B$ and $\mu_n$, pionic effects are small and only visible
around $\rhoB \sim 0.4~\fm^{-3}$,
while we find large differences in $\mu_p$.
The equality $\mu_C=\mu_p-\mu_n=-m_\pi$ under $\pi^-$ condensation
reads the Fermi energy relation, $E_F(n)=E_F(p)+m_\pi$.
Since a neutron on the Fermi surface is replaced with a proton and a $\pi^-$
having the same total energy,
we have to pay the cost of the pion rest mass energy
in exchange for the Fermi energy reduction and symmetry energy gain.
At higher densities where hyperons appear, 
pionic effects becomes smaller, and disappear at $\rhoB=0.88~\fm^{-3}$.

The $s$-wave pion condensation would be suppressed
when we include the $\pi{N}$ interaction.
We evaluate the pion energy
by using the potential of the form~\cite{PionicAtom,Batty1983,Kienle,Friedman}, 
\begin{equation}
U_s(\pi^-)=-\frac{2\pi}{m_\pi}\left[
	\left(1+\frac{m_\pi}{M_N}\right)\left(b_0\rhoB
	+b_1\delta\rho\right)
	+\left(1+\frac{m_\pi}{2M_N}\right)\mathrm{Re}B_0\,\rhoB^2
	\right]
	\ ,
\end{equation}
where $b_1=b_1^\mathrm{free}/(1-\alpha\rhoB/\rho_0), \delta\rho=\rho_n-\rho_p$.
As typical examples,
we adopt the parameter sets
from the analyses of pionic atom data;
$b_0=-0.023/m_\pi$, $b_1=-0.085/m_\pi$ ($\alpha=0$),
$\mathrm{Re}B_0=-0.021/m_\pi^4$ (less repulsive)~\cite{Batty1983},
and
$b_0=-0.0233/m_\pi$, $b_1^\mathrm{free}=-0.1473/m_\pi$, $\alpha=0.367$
($b_1=-0.1149/m_\pi$ at $\rhoB=0.6\rho_0$),
$\mathrm{Re}B_0=-0.019/m_\pi^4$ (more repulsive)~\cite{Kienle}.
In Fig.~\ref{Fig:Pot}, 
we show the pion energy, $E_\pi=\sqrt{m_\pi^2+2m_{\pi}U_s}$,
calculated with these potentials and proton fraction in TM1 EOS.
With these potentials,
we find that the existence of the $s$-wave pion condensed region,
$E_\pi < \mu_e$,
depends on the pion optical potential parameters.
Since the pion potential above the normal nuclear density is not yet known,
the realization of pion condensation may be marginal and model-dependent.

\begin{figure}
\centerline{
\Psfig{12cm}{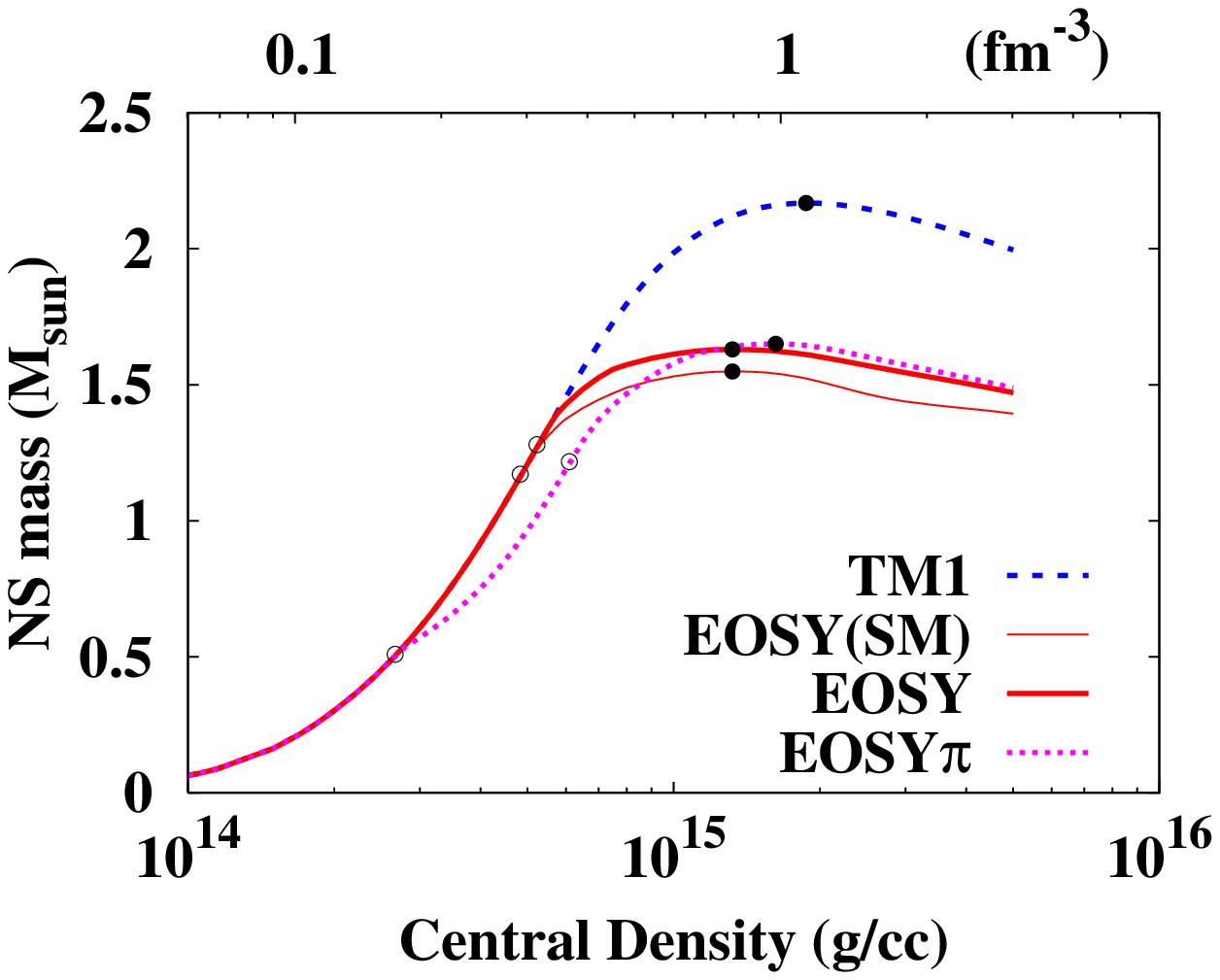}
}
\caption{Neutron star masses are shown as functions of central density.
The dashed, thin solid, thick solid, dotted lines show
the results with
nucleonic EOS (TM1~{\protect\cite{TM1}}), 
hyperonic EOS with attractive hyperon potentials
(\EOSY(SM)~{\protect\cite{Schaffner}}),
hyperonic EOS with repulsive hyperon potentials (\EOSY)
and hyperonic EOS with pions (\EOSYpi),
respectively.
Filled points show the maximum neutron star masses,
and open circles show the threshold densities of hyperons and pions.
}
\label{Fig:NS}
\end{figure}

We apply the above four EOSs of neutron star matter discussed above 
(TM1 EOS, \EOSY(SM), \EOSY, \EOSYpi)
to the hydrostatic structure of neutron stars 
by solving the Tolman-Oppenheimer-Volkoff equation.  
We plot the gravitational mass of neutron stars 
as a function of central baryon mass density in Fig.\ref{Fig:NS}.  
The maximum mass of neutron stars in \EOSY\ 
is smaller than the case of nucleonic EOS
because of the softness from hyperons.
The maximum mass is $1.63~M_\odot$ for \EOSY\ 
in contrast to $2.17~M_\odot$ for nucleonic EOS
when we adopt repulsive potentials for hyperons.
The maximum mass is further reduced 
to be $1.55~M_\odot$ in \EOSY(SM) with attractive potentials.
The neutron star masses with \EOSYpi\  
are reduced in the mid range of central density
$0.16~\fm^{-3}<\rhoB<0.8~\fm^{-3}$,
but the maximum mass ($1.65 M_\odot$) is almost the same.
This is because the maximum mass is mainly determined
by the EOS at densities around $0.8~\fm^{-3}$ or more,
where the condensed pion density is small.
The central density of a typical neutron star
having 1.4$M_\odot$ is 0.35 fm$^{-3}$ in nucleonic EOS (TM1),
which is a little above the threshold density of $\Lambda$ in \EOSY.
In this case, hyperons are limited only in the core region,
and the neutron star mass does not get a large reduction
as seen in Fig.~\ref{Fig:NS}.  
We summarize the neutron star masses and the threshold densities
to have hyperons in neutron star matter in Table \ref{Table:NS}.

\subsection{Hyperonic matter at finite temperatures}

We next study the EOS of supernova matter,
where the hadronic charge fraction ($Y_C$) is fixed at finite temperature.
We here show the results including electron and photon contributions.
Finite nuclear effects in Shen EOS are included.
The treatment of leptons in supernova matter are explained in Appendix A.4.

\begin{figure}
\centerline{
\Psfig{16cm}{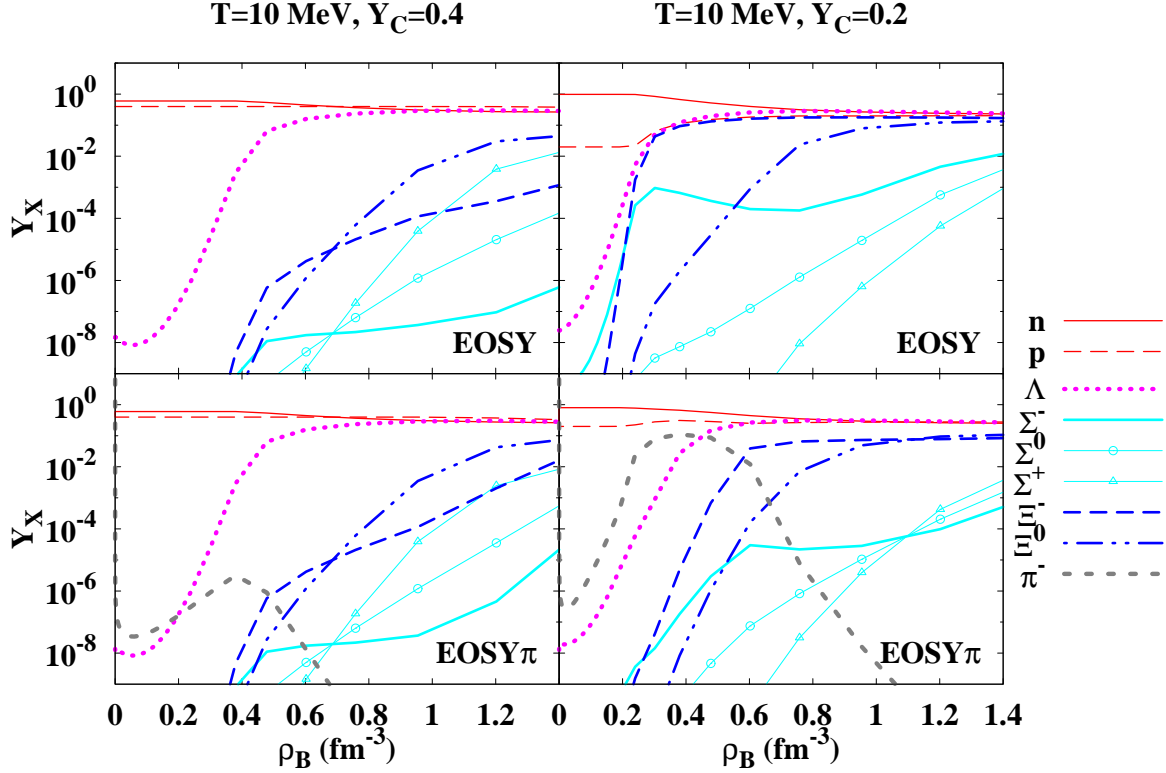}
}
\caption{Composition of supernova matter
at  $(T,Y_C)=(10~\mathrm{MeV}, 0.4)$ (left)
and $(T,Y_C)=(10~\mathrm{MeV}, 0.2)$ (right)
in the hyperonic EOS table without (\EOSY, upper)
and with (\EOSYpi, lower) pions.  
The number fraction of particles are plotted as functions 
of baryon density.  The species of particles are denoted 
as in the legend.}
\label{Fig:Const-SN}
\end{figure}

In order to demonstrate the contents of the EOS table,
we show the energy and compositions as functions of 
baryon density by choosing $T=10~\MeV$ and $Y_C=0.4$
as an example.
In the upper-right panel of Fig.~\ref{Fig:E/B}, 
we plot energy per baryon ($E/B$) in \EOSY\ 
together with the results in Shen EOS for comparison.
At $\rhoB > 0.4~\fm^{-3}$, 
the energy is lower in \EOSY\ than in Shen EOS due to hyperons
(See Fig.~\ref{Fig:Const-SN}).
In the lower-right panel of Fig.~\ref{Fig:E/B},
neutron and proton chemical potentials are shown.  
In \EOSY, the difference of chemical potentials 
between neutron and proton is small and 
neutron chemical potential becomes smaller 
than proton chemical potential at 0.7 fm$^{-3}$.
This is because more $\SU_f(3)$ symmetric matter is preferred
toward high densities,
and the charge fraction under $\beta$ equilibrium
decreases below $Y_C=0.4$ at high densities.

We show the particle compositions as functions of baryon density
in \EOSY\ and \EOSYpi\ at $Y_C=0.4$ in the left panels
of Fig.~\ref{Fig:Const-SN}.
In moderately isospin asymmetric matter ($Y_C=0.4$),
the fraction of $\Lambda$ particle grows at around $\rhoB \sim 0.4~\fm^{-3}$, 
and becomes comparable to nucleons at higher densities.
The fractions of other strange baryons increase slowly 
and remains small until very high density.
In the right panels of Fig.~\ref{Fig:Const-SN},
we show the particle compositions at $Y_C=0.2$.
With this small charge fraction,
total hyperon fraction reaches 1 \% at $0.25~\fm^{-3}$.

\begin{figure}
\centerline{
\Psfig{13cm}{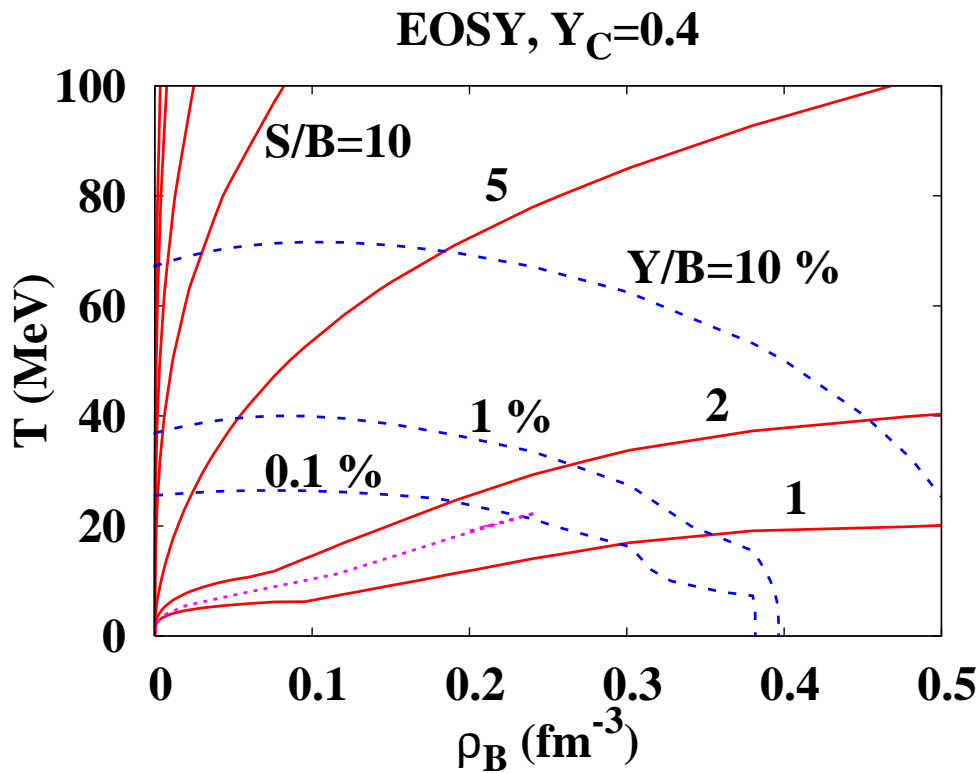}
}
\caption{
Hyperon fraction contours and adiabatic paths
in supernova matter at $Y_C=0.4$
from the hyperonic EOS table without pions (\EOSY). 
The contours of the fixed number fraction of hyperons 
(sum of strange baryons) are shown by dashed lines.  
The solid lines denote the contour of fixed entropy 
per baryon (isentropy).  
The dotted line shows the trajectory of the dense 
matter at center during core collapse and bounce.}
\label{Fig:ADP}
\end{figure}

In Fig.~\ref{Fig:ADP}, we plot the contour map of 
the fraction of hyperons (sum of strange baryons) 
in the density-temperature plane.  
In order to have a significant amount of hyperons, 
one needs high density or temperature.  
In supernova core, the entropy per baryon is typically 
around 1--2 k$_B$, therefore, one needs high densities 
0.3--0.4 fm$^{-3}$ to have 1$\%$ mixture of hyperons and 
0.45 fm$^{-3}$ for 10$\%$.

In supernova matter, the effects of hyperons and pions are limited.
When isospin asymmetry is not high (ex. $Y_C = Y_e = 0.4$),
total amount of hyperons which appears at $0.4~\fm^{-3}$
is around 1\% of baryons.
The value of $Y_C$ remains high due to the neutrino trapping
during the collapse and bounce~\cite{neutrino}.
However, after the deleptonization
due to neutrino emission, $Y_C$ becomes smaller and hyperons may appear
in the proto-neutron star cooling process.
Dense matter at higher densities and temperatures
may appear also in black hole formations,
and hyperon effects can be expected in such processes.

\subsection{Applications to core-collapse supernovae}

As an application of the EOS table with hyperons, 
we perform the numerical calculations of hydrodynamics 
of core-collapse supernovae.  
This calculation is aimed to test the data of EOS table 
for numerical simulations and to provide the basic 
information on the properties of EOS in supernovae 
such as the appearance of hyperons.  
For this purpose, we calculate the adiabatic collapse of 
iron core of massive stars of 15$M_\odot$~\cite{Woosley}.
In the same way as the hydrodynamical calculations
in Sumiyoshi et al.~\cite{Sumiyoshi2004},
we calculate the general relativistic hydrodynamics 
under the spherical symmetry without neutrino-transfer, 
which is time-consuming, 
by assuming that the electron fraction is fixed 
to the initial value in the stellar model.
Numerical simulations by neutrino-radiation hydrodynamics 
are in progress.  

We have found that the adiabatic collapse of 15$M_\odot$ star 
with \EOSY\ leads to a prompt explosion.  
This {\it model} explosion is caused by the large electron fraction 
assumed and is quite similar to the case obtained
with Shen EOS~\cite{Sumiyoshi2004}.
The explosion energy is almost the same as the case with Shen EOS 
and the difference turns out to be small within 0.5\%.  
We plot the trajectory of density and temperature 
of the central grid in the hydrodynamical calculation
in Fig.~\ref{Fig:ADP}.  

We have examined the appearance of hyperons during the evolution 
of core-collapse and bounce.  
We find that the fraction of hyperons turn out to be 
very small within 10$^{-3}$.  
This is because the density does not increase drastically 
even at the core bounce in the current model.
The peak density is 0.24 fm$^{-3}$ which is lower than 
the threshold density 0.60 fm$^{-3}$ at temperature 21.5 MeV
and electron fraction 0.42, where $\Lambda$ hyperons appear
by the same order as nucleons.  
This small mixture does not affect largely the dynamics 
in the model explosion.

A large electron fraction leads to a large proton fraction, 
and, therefore, suppresses the appearance of hyperons~\cite{neutrino}.
We note here that this is the outcome of simple adiabatic 
hydrodynamics without the treatment of neutrinos.  
When electron captures and neutrino trapping are taken 
into account~\cite{Sumiyoshi2005},
electron fraction might be smaller than the current value
and may enhance the hyperon appearance.  
The hyperons will definitely appear in the thermal evolution 
of proto-neutron stars after 20 seconds~\cite{Pons},
during which the central 
density becomes high and the electron fraction gets smaller.  
In recent findings of black hole formation from massive stars 
of 40$M_\odot$~\cite{Sumiyoshi2006,Sumiyoshi2007}, the hyperon EOS is necessary 
since the density becomes extremely high during the 
collapse toward the black hole.  
It would be interesting to perform the full simulations 
of core-collapse supernovae and related astrophysical 
phenomena.  

\section{Summary and discussion}
\label{Sec:Summary}

In this paper, 
we have presented several sets of equation of state (EOS) of supernova matter
(finite temperature nuclear matter with lepton mixture)
including hyperons (\EOSY)
using an $\mathrm{SU}_f(3)$ extended relativistic mean field (RMF) model
with a wide coverage of density, temperature, and charge fraction.
Supernova matter EOS is one of the most essential parts
in numerical simulations of core collapse supernovae.
At present, two sets of supernova matter EOS 
(Lattimer-Swesty EOS~\cite{LS-EOS} and Shen EOS~\cite{Shen-EOS})
are widely used.
The constituents in these EOSs are nucleons and nuclei,
then it is desired to include hyperons,
which are believed to appear at high densities.
Here we have extended the relativistic EOS by Shen et al.~\cite{Shen-EOS}
by introducing hyperons.

We start from the RMF parameter set TM1 for nucleon sector~\cite{TM1},
which well describes the bulk properties of nuclei
in the wide mass and isospin range.
For hyperon-meson coupling constants,
we adopt the values in Ref.~\cite{Schaffner} as the starting points.
Hyperon-vector meson couplings are fixed
based on the flavor-spin SU(6) symmetry, 
and hyperon-scalar meson couplings are determined
to give the hyperon potentials in nucleonic and hyperonic matter.
Hyperon potentials in nuclear matter around the normal density,
$U^{(N)}_Y(\rho_0)$, 
are accessible in hypernuclear production reactions.
Recent developments in hypernuclear physics suggest 
the following potentials for
$\Sigma$~\cite{Harada,Kohno,Maekawa}
and $\Xi$
baryons~\cite{Twin,Fukuda,HYP06-Maekawa,MaekawaXi}
\begin{equation}
U^{(N)}_\Sigma(\rho_0)\simeq +30~\mathrm{MeV}\ ,\quad
U^{(N)}_\Xi(\rho_0)\simeq -15~\mathrm{MeV}\ .\quad
\end{equation}
These potentials are consistent with those in the quark-cluster model
for $YN$ interaction~\cite{KohnoQC}
and a chiral model prediction~\cite{Kaiser}.
In this paper, we have modified $g_{\sigma\Sigma}$ and $g_{\sigma\Xi}$
to explain these potentials,
while other coupling constants are unchanged
from those in Ref.~\cite{Schaffner}.

The $\Sigma$ potential in nuclear matter still has ambiguities.
Recent theoretical analysis~\cite{Maekawa} has shown that 
the shape and absolute values of quasi-free $\Sigma$ production spectra
are well explained in a Woods-Saxon potential
with $U_\Sigma(\rho_0) \simeq +15~\mathrm{MeV}$.
On the other hand, a few MeV attractive pocket is known to be required
to explain the energy shift of $\Sigma^-$ atom~\cite{Batty,Mares,HYP06-Tsubaki},
then the central repulsion would be stronger
to cancel the effects of this pocket.
Other theoretical analyses~\cite{Harada,Kohno} suggest that 
$U_\Sigma(\rho_0) \simeq +30~\mathrm{MeV}$ would be preferred
in order to explain the shape or the absolute yield
in $\Sigma$ production spectra.
In any of these analyses,
$\Sigma$ potential should be repulsive or less attractive
than that for $\Lambda$,
then the effects of $\Sigma$ hyperons are much smaller
than those in the attractive case.
It is to be noted that
the ambiguities in $U_\Sigma$ do not affect the supernova EOS very much
as far as $\Sigma$ hyperon fraction is small.

Formation of finite nuclei at low densities 
is another important ingredient in supernova simulations.
In the present EOS, 
effects of finite nuclear formation are included
by using the Shen EOS~\cite{Shen-EOS},
in which formation of finite nuclei is included
in the Thomas-Fermi approximation.
Effects from finite nuclei are evaluated
by the difference of free energy and its derivatives
in the Shen EOS from the EOS of uniform nucleonic matter (TM1) without hyperons 
at each $(T, \rhoB, Y_C)$.

We have examined the properties of the EOS with hyperons
in neutron star matter ($T=0$, $\beta$-equilibrium)
and supernova matter.
Hyperon effects are significant in neutron stars
as discussed already in the literature~\cite{Glendenning,Schaffner,Balberg,HYP00-Sahu,HYPMAT,Sugahara1994,Shen2002}.
Hyperons appear at around $\rhoB \simeq 2 \rho_0$
in cold matter under $\beta$-equilibrium and soften the EOS.
The maximum mass of neutron stars decreases
from $2.17 M_\odot$ 
to $1.55 M_\odot$ and $1.63 M_\odot$
when hyperons are included with attractive and repulsive hyperon potentials,
respectively.
In prompt phase in supernova explosions, on the other hand,
hyperon effects are found to be small
in a spherical, adiabatic collapse of a 15$M_\odot$ star by the hydrodynamics
without neutrino transfer.
In the case with $Y_C=Y_e=0.4$ as a typical example, 
hyperon fraction becomes meaningful ($Y_Y>1 \%$)
at $\rhoB>0.4~\fm^{-3}$ or $T>40~\MeV$.
In the spherical and adiabatic core collapse calculation
of a massive star with the 15$M_\odot$~\cite{Woosley},
the maximum density and temperature are found to be
$(\rhoB,T)=(0.24~\fm^{-3}, 22~\MeV)$, 
which do not reach the region of the above hyperon mixture region.
It should be noted that this conclusion is model dependent.
Hyperons may appear more abundantly in more realistic calculations
with neutrino transfer, which are in progress.

We have also discussed the roles of pions in neutron stars and supernovae.
In this work, we have examined the effects of
free thermal pions~\cite{Glendenning}.
In neutron star matter, the absolute value of the charge chemical potential
$\mu_C=\mu_p-\mu_n$
is calculated to be larger than the pion mass at $\rhoB \gtrsim \rho_0$,
thus charged pions can condensate
as far as the pion-nucleon interaction is not very repulsive.
The EOS softening from pions is moderate and limited in the density range
$\rhoB<0.88~\fm^{-3}$ without ($p$-wave) $\pi N$ attraction,
then the maximum mass of neutron stars ($1.65 M_\odot$)
is almost the same as that without pions.
In supernova explosions,
temperatures are not very high and pion contributions are small.
At higher temperatures as in the case of black hole formation
or high energy heavy-ion collisions, the role of pions should be significant.

There are several points to be improved
for deeper understanding of supernova matter EOS.
First, it is necessary to examine the coupling constants
of hyperons with hidden strangeness mesons, 
$\zeta$ and $\phi$, which critically decide $YY$ interaction.
In this paper, we have adopted $g_{\zeta Y}$ and $g_{\phi Y}$
in Ref.~\cite{Schaffner},
where the couplings are determined based on the SU(6) relation
and a conjecture on the hyperon potential depth in hyperon matter.
An alternative way to determine these couplings would be to invoke
various hypernuclear and hyperon atom data, such as
the double $\Lambda$ hypernuclear binding energy
in $^6_{\Lambda\Lambda}\mathrm{He}$~\cite{Nagara}
and atomic energy shifts in $\Sigma^-$ atom~\cite{Batty,Mares,HYP06-Tsubaki}.
In Ref.~\cite{HYP06-Tsubaki}, 
Tsubakihara et al. have determined the scalar couplings of
$g_{\sigma\Lambda}$, $g_{\zeta\Lambda}$,
$g_{\sigma\Sigma}$ and $g_{\zeta\Sigma}$,
by using the double $\Lambda$ hypernuclear bond energy
and the atomic energy shift of $\Sigma^-$ atom,
while vector couplings are fixed from the $\SU_f(3)$ relations.
At present, available data are so scarce that we cannot fix these couplings
based on the data unambiguously, but future coming J-PARC and FAIR facilities
will provide much more data on $YY$ interaction.
Next, it is desired to respect chiral symmetry
in order to describe very dense matter, in which spontaneous broken
chiral symmetry will be partially restored.
A chiral symmetric RMF model~\cite{Tsubaki2007} is recently developed
based on a scalar meson self-energy
derived in the strong coupling limit of lattice QCD,
and it describes binding energies and radii of normal nuclei
in a comparable precision to TM1.
An $\SU_f(3)$ extended version of this chiral RMF
is now being developed~\cite{HYP06-Tsubaki}.
Finally, distribution of finite nuclear species may be important
at low densities~\cite{Ishizuka}.
At finite temperatures,
the entropy increase by the formation of various fragments
will contribute to gain the free energy
compared with the single heavy-nuclear configuration
assumed in the Thomas-Fermi approach.
It is not straightforward but challenging
to include nuclear statistical equilibrium (NSE) distribution
in a consistent way in the EOS based on RMF.

These challenging developments of hadronic and nuclear physics
are important to understand the extreme conditions in compact
objects and to clarify the mechanism of explosive phenomena
in astrophysics.

\section*{Acknowledgements}

We would like to thank Dr. Daisuke Jido and Mr. Takayasu Sekihara
for useful discussions.
This work is supported in part by the Ministry of Education,
Science, Sports and Culture,
Grant-in-Aid for Scientific Research
under the grant numbers,
    15540243,	
    1707005,	
    18540291,
    18540295
and 19540252,	
the 21st-Century COE Program "Holistic Research and Education
Center for Physics of Self-organization Systems",
and Yukawa International Program for Quark-hadron Sciences (YIPQS).
The authors also would like to thank 
National Astronomical Observatory of Japan (NAOJ),
Japan Atomic Energy Agency (JAEA),
and 
Yukawa Institute for Theoretical Physics (YITP)
for computing resources.

\appendix
\section{Note on the EOS table}

\subsection{Locations of data tables}
The data tables are available on \\
{\verb http://nucl.sci.hokudai.ac.jp/~chikako/EOS/index.html }\\
or upon request to A. Ohnishi.
On the web page,
the tables under the name of '***.tbl' (EOS table)
and '***.rat' (composition table) are available
for the sets using $U_\Sigma^{(N)}=-30, 0, +30, +90~[\MeV]$
at normal density.
The most recommended potential is $U_\Sigma^{(N)}=+30~\MeV$.
For other hyperons,
we adopt the potential depth $U_\Lambda^{(N)}=-30~\MeV$ 
and $U_\Xi^{(N)}=-15~\MeV$
as described in section \ref{Subsec:RMFY}.
The EOS table with thermal pions are available in addition
to the standard choice without pions.
As described in section \ref{Subsec:pions},
the set with thermal free pions
are aimed only for the assessment of pion contributions
in a simple treatment.
Further careful treatment of pion interactions is necessary.

\subsection{Definition of quantities in the EOS table}
We list the definitions of the physical quantities tabulated in the
EOS table, '***.tbl'.
We note that the order of quantities in the list is partly different
from the original Shen EOS table \cite{Shen-EOS}
since the current table contains lepton and photon contributions.
The definitions of quantities follows the ones in Shen EOS
unless stated specifically below.

\begin{itemize}
\item[(1)] Logarithm of baryon mass density:
$\log_{10}(\rhoB)~[g/\mathrm{cm}^3]$ \\
The baryon mass density $\rhoB$ is defined by
\begin{equation}
\rhoB = M_u n_B
\end{equation}
where $M_u$ and $n_B$ are the atomic mass unit and the baryon number density,
respectively.

\item[(2)] Charge ratio: $Y_CC= n_C/n_B$
The charge density $n_C$ is defined by
\begin{equation}
n_C= \displaystyle{\sum_B}q_i n_i
\ ,
\end{equation}
where $q_i$ and $n_i$ are the charge and the number density
of the baryons and the sum runs over the baryon octet.

\item[(3)] Entropy per baryon: $S/B~[k_B]$ \\
The entropy per baryon
contains the contributions from hadrons, leptons and photons.

\item[(4)] Temperature: $T~[\MeV]$ \\

\item[(5)] Pressure: $P~[\MeV/\fm^{3}]$ \\
The pressure
contains the contributions from hadrons, leptons and photons.

\item[(6)] Chemical potential of neutron: $\mu_n~[\MeV]$ \\
The chemical potential of neutron is measured
relative to the nucleon mass $M_N = 938~\MeV$.
It is connected with the baryon chemical potential as
\begin{equation}
\mu_n=\mu_B - M_N
\end{equation}

\item[(7)] Chemical potential of proton: $\mu_p~[\MeV]$ \\
The chemical potential of  proton is measured
relative to the nucleon mass $M_N$.
The relation to the baryon and charge chemical potentials reads
\begin{equation}
\mu_p=\mu_B + \mu_C - M_N
\end{equation}

\item[(8)] Chemical potential of electron:  $\mu_e~[\MeV]$ \\
The chemical potential of electron is determined by the charge
neutrality as
\begin{equation}
n_e = \displaystyle{\sum_B}q_i n_i.
\end{equation}
See below for the descriptions on the lepton contributions.

\item[(9)] Free neutron fraction: $Y_n$  \\
In the uniform matter, the free neutron fraction is simply
the ratio, $n_n/n_B$.  For the non-uniform matter at low density,
the definition follows the one in Shen EOS.

\item[(10)] Free proton fraction: $Y_p$  \\
The ratio, $n_p/n_B$, as in $Y_n$ above.
For the fraction of strangeness baryons,
see the description below on the table of number fractions.

\item[(11)] Mass number of heavy nucleus: $A$   \\
The values of from $(11)$ to $(14)$ are taken from Shen EOS table
or zero set above normal nuclear density.

\item[(12)] Charge number of heavy nucleus: $Z$

\item[(13)] Heavy nucleus fraction: $X_A$

\item[(14)] Alpha-particle fraction: $X_{\alpha}$

\item[(15)] Energy per baryon: $E/B~[\MeV]$ \\
The energy per baryon
is defined with respect to the free nucleon mass $M_N$ and
contains the contributions from hadrons, leptons and photons.

\item[(16)] Free energy per baryon: $F/B~[\MeV]$ \\
The free energy per baryon
is defined with respect to the atomic mass unit $M_u$ and
contains the contributions from hadrons, leptons and photons.

\item[(17)] Effective mass: $M^*~[\MeV]$ \\
The effective mass of nucleon is obtained in the RMF theory
for uniform matter.
In non-uniform matter,
we replace the effective mass $M_N^*$ by the free nucleon mass $M_N$.
\end{itemize}

\subsection{Data table of composition}

In order to provide the information on the appearance of hyperons
other than nucleons, we prepare a separate data table (***.rat)
for the number fractions.
The number fraction, $Y_i = n_i / n_B$, is given
as a function of $(\rhoB, T, Y_C)$ in the following order.


\begin{itemize}
\item[(1)] Logarithm of baryon mass density:
	$\log_{10}\rhoB~[\mathrm{g}/\mathrm{cm}^3]$
\item[(2)] Temperature: $T~[\MeV]$
\item[(3)] Charge ratio: $Y_C$
\item[(4)] Neutron ratio (including neutrons in alpha): $Y_n$
\item[(5)] Proton ratio (including protons in alpha): $Y_p$
\item[(6)] $\Lambda$ ratio: $Y_{\Lambda}$
\item[(7)] $\Sigma^-$ ratio: $Y_{\Sigma^-}$
\item[(8)] $\Sigma^0$ ratio: $Y_{\Sigma^0}$
\item[(9)] $\Sigma^+$ ratio: $Y_{\Sigma^+}$
\item[(10)] $\Xi^-$ ratio: $Y_{\Xi^-}$
\item[(11)] $\Xi^0$ ratio: $Y_{\Xi^0}$
\end{itemize}

\subsection{Treatment of leptons}
We describe briefly on the contribution of leptons 
(electrons, muons and neutrinos) in the current study.  

We remark that we take into account muons 
in the case of cold neutron stars.  
Muons appear abundantly at densities higher than the threshold 
when the electron chemical potential exceeds the muon rest mass.  
The chemical potentials of electrons and muons are related with 
\begin{equation}
\mu_\mu = \mu_e
\end{equation}
This is because neutrinos freely escape from the neutron star 
and are not trapped inside.
Accordingly, the contributions of electrons and muons are 
taken into account in the discussions of cold neutron stars.

For the EOS table for core-collapse supernovae,
we add the contributions of electrons, 
positrons and photons while muons are not added because of 
the following reason.  
In the supernova cores, 
neutrinos are trapped inside the supernova core 
and the Fermi energies of neutrinos becomes high and non-zero.
If the chemical equilibrium holds, the chemical potentials follows 
the relations,
\begin{equation}
\mu_\mu - \mu_{\nu_\mu} = \mu_e - \mu_{\nu_e}
\ ,
\end{equation}
\begin{equation}
n_\mu + n_{\nu_\mu} = 0
\ .
\end{equation}
The latter relation comes from the fact that 
the net $\mu$-type lepton number is zero.  
Because of positive values of $\mu_{\nu_e}$, 
r.h.s of the chemical equilibrium is reduced.  
In addition, the appearance of muon requires 
the production of anti-neutrinos of $\mu$-type 
and leads to the negative value of $\mu_{\nu_\mu}$.  
Therefore, the appearance of muons are suppressed 
in supernova core.  

Contributions of neutrinos are not added 
because the treatment depends on the density region.  
In the central part of supernova core, 
neutrinos are trapped and the chemical equilibrium 
are reached together with neutrinos.
Outside the neutrino trapping surface,
typically at $\sim$ 10$^{11}$ g/cm$^{3}$,
neutrinos escape freely and neutrinos do not contribute.  
They also depend on 
the method of neutrino-radiation in numerical simulations.  


\end{document}